# Intermolecular interaction enhances thermoelectric performance of molecular junctions

Abdalghani Daaoub[1], Renhang Wang[1], Ross Davidson[2], Sara Sangtarash[1], Hatef Sadeghi[1,*]

[1] Device Modelling Group, School of Engineering, University of Warwick, Coventry CV4 7AL, United Kingdom
[2] Chemistry Department, Durham University, Durham DH1 3LE, United Kingdom
* *Hatef.Sadeghi@warwick.ac.uk*

**Abstract**
Novel organic materials formed from functional molecules are attractive for various nanoelectronic applications because they are environmentally friendly, widely available and inexpensive. Recent advancement in bottom-up fabrication methods has made it possible to design and synthesis functional molecules with desired functionalities and engineer their properties precisely. This requires deeper understanding of if the properties of building blocks e.g. single molecules can be translated to many molecule junctions in the form of self-assembled monolayers (SAM). Therefore, understanding the effect of intermolecular interaction becomes important. In this paper, we study the effect of intermolecular interactions on the charge transport and thermoelectric properties of junctions formed by parallel molecules between metallic electrodes. We demonstrate that the electrical conductance and Seebeck coefficient are enhanced simultaneously leading to more than an order of magnitude enhancement of power factor as a result of intermolecular coupling between two molecules of the same kind or different. This strategy can be used to improve the thermoelectric performance of SAMs by engineering their packing density.

**Introduction**
Organic molecular scale thermoelectricity has attracted increasing interest over the past years because they can offer environmental friendly and inexpensive solution that is not accessible in conventional inorganic thermoelectric materials[1–4]. Furthermore, the thermoelectric and electronic properties of molecules can be tuned using synthetic chemistry approaches[2,5,6]. For example, by only changing the anchor group[7–9] or metallic centre[10] or addition of a heteroatom[11], the sign of thermopower (Seebeck coefficient) can be tuned in molecules. This is very useful because for any thermoelectric device, *n*-type and *p*-type materials should be arranged in a tandem device configuration to enhance the power output[12].

The field of molecular scale thermoelectricity relatively new and is still in its development phase[1,2,13–15]. The focus is mainly on developing new strategies to enhance their efficiency. The efficiency of a thermoelectric material is characterised by the dimensionless thermoelectric figure of merit[16] $ZT=PT/\kappa$ where $P=GS^2$ is power factor, $G$ is electrical conductance, $S$ is Seebeck coefficient, $\kappa$ is thermal conductance and $T$ is temperature. Clearly, $P$ needs to be increased while $\kappa$ should be decreased to improve $ZT$. $ZT$ values higher than 3 will lead to economically viable technology[1]. Improvement of $ZT$ is constrained by the interdependence of thermoelectric parameters[17]. For example, both $G$ and $S$ depend on the transmission probability of electrons passing through a thermoelectric material. In most molecules studied today, increase in $G$ leads to decrease of $S$ and vice versa[13].

Table 1 shows the power factor of some of the molecules with the highest measured $G$ or $S$ studied so far. Clearly, when $S$ increases, $G$ decreases in molecular junctions with similar backbone. For example, when the length of the molecules with the thiophene core is increased from one thiophene unit to 3, 7-fold enhancement of $S$ is achieved while $G$ decreases by a factor of 3. Ideally, $S$ and $G$ need to be enhanced simultaneously to improve the efficiency of thermoelectric materials[18,19].

In this paper, we demonstrate a generic strategy to enhance the efficiency of molecular materials by simultaneous enhancement of $G$ and $S$. We examine the effect of intermolecular interactions on thermoelectric properties and demonstrate for the first time that simultaneous enhancement of $G$ and $S$ is achievable in molecular complexes formed by electron deficient benzo[1,2-c;4,5-c′]bis[1,2,5]thiadiazole (BBT)[20] and anthracene[21–23] molecules with similar length. We show that the power factor can be enhanced by more than an order of magnitude. This shows that the density of self-assembled monolayers (SAMs) formed by such molecular structures and the intermolecular interactions play important role in the thermoelectric properties of molecular materials and can be used to further enhance their performance.

**Table 1.** Examples of conductance $G$, Seebeck coefficient $S$ and power factor $P$ values measured for some single molecular junctions.

| Molecule | Molecule structure | $G(G_0)$ | $S\ (\mu V/K)$ | $P(pW/K^2)$ | Ref |
|---|---|---|---|---|---|
| 4,7-Dithiophenyl-2,1,3-benzothiadiazole-3,3-dithiol (DTBTDT) | | $9.2 \times 10^{-3}$ | $15.46 \pm 0.15$ | $1.70 \times 10^{-4}$ | 24 |
| 1,4-Butanedithiol (C4) | | $4.5 \times 10^{-3}$ | $2.1 \pm 0.11$ | $1.53 \times 10^{-6}$ | 24 |
| 1,6-Hexanedithiol (C6) | | $6.2 \times 10^{-4}$ | $5.55 \pm 0.13$ | $1.47 \times 10^{-6}$ | 24 |
| S,S′-2,2′-(Thiophene-2,5-diyl)bis(ethane-2,1-dyil)-diethanethioate (TA3) | | $2 \times 10^{-3}$ | 2.2 | $7.49 \times 10^{-7}$ | 25 |
| S,S′-([2,2:5′,2′′-Terthiophene]-5,5′′-diyl)-diethanethioate (OT3) | | $7 \times 10^{-4}$ | 14.84 | $1.19 \times 10^{-5}$ | 25 |
| Oligomers of thiophene-1,1-dioxide (TDOn, n = 1, 2, 3, 4) | | $\sim 9 \times 10^{-4}$<br>$\sim 5 \times 10^{-4}$<br>$\sim 2 \times 10^{-4}$<br>$\sim 6 \times 10^{-5}$ | 7.3<br>6.4<br>2.4<br>−22.1 | $3.71 \times 10^{-6}$<br>$1.58 \times 10^{-6}$<br>$8.92 \times 10^{-8}$<br>$2.27 \times 10^{-6}$ | 26 |
| C60 | | 0.1 | $-18 \pm 6.8$ | $2.5 \times 10^{-3}$ | 27 |
| Stable Blatter radical | | $3.16 \times 10^{-4}$ | $25.9 \pm 2.1$ | 0.21 | 28 |

**Result and discussion**

Figure 1a shows the schematic of junctions including molecules **1** and **2** with BBT and anthracene cores (Figure 1b,c) connected to the thiol anchor groups through acetylene linkers. We first calculate the ground-state geometry and electronic structure of gas phase molecules **1** and **2** using SIESTA[29] implementation of density functional theory (DFT) as discussed in the method section. Although the frontier orbitals for these two molecules are similar, the energy gap and the position of the highest occupied molecular orbital (HOMO) and the lowest unoccupied molecular orbital (LUMO) energies and the HOMO-LUMO energy gap are different. Our calculation shows that the HOMO-LUMO gap of **1** is 110 meV smaller than **2**. Similar behaviour was obtained using three different DFT functionals (See supporting information and computational methods for details).

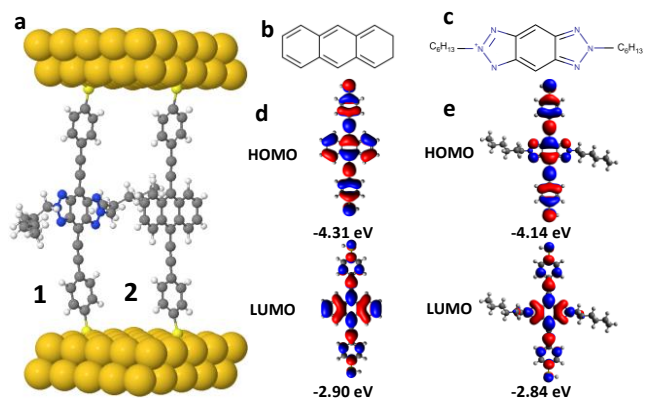

**Figure 1. Chemical structures of the molecular structures.** (a) Molecules **1** and **2** between gold electrodes with anthracene (b) and BBT (c) molecular cores. (d, e) Frontier orbitals of molecules **1** and **2** with BBT and anthracene core, respectively and the thioacetyl (SAc) anchor groups. The molecular length of molecules **1** and **2** is the same.

Next, we calculate transport properties of molecules **1** and **2** between electrodes using two gold electrodes with the flat surfaces (Figure 1a). From the ground-state relaxed geometry of molecules between electrodes, we calculate the mean-field Hamiltonian of junctions using DFT and combine it with our transport code GOLLUM[16,30] to calculate the transmission coefficient $T(E)$ of electrons with energy $E$ passing from one electrode to the other through molecules **1** and **2** (structures A and B in Figure 2a). Figure 2b shows $T(E)$ for the single molecules **1** and **2** between electrodes (green and red lines, respectively). The DFT Fermi energy is closer to the HOMO resonances for both molecules which is usually expected for molecules with the thiol anchors[31]. The transport HOMO-LUMO gap is smaller for **1** in agreement with energy gaps calculated for gas phase molecules. Smaller HOMO-LUMO gap leads to higher $T(E)$ for **1** compared to that of **2** for all energies between the HOMO-LUMO.

We now study the effect of intermolecular interactions between molecules **1** and **2** on $T(E)$. For this we first calculate the optimum distance between molecules **1** and **2** in gas phase and find it to be 3.35Å. We then place the two molecules between electrodes and perform geometry optimisation and found that Au-S relaxes to hollow conformation (structure C in Figure 1a). Orange curve in Figure 2b shows $T(E)$ of configuration C. First, the HOMO-LUMO gap of two molecules in parallel (structure C) is smaller than both molecules **1** (structure B) and **2** (structure A). Secondly, the transmission is higher for C compared to that of A and B. This is higher than what is expected classically as shown in Figure 2c where the amplitude of orange curve ($T(E)$ of C) is higher than that of the black dashed curve ($T(E)$ of A+B). We attribute this to the intermolecular interaction between the molecules.

To demonstrate that this is due to intermolecular interaction, we increase the distance between the molecules (structure D and E in Figure 2a) and calculate $T(E)$. As the distance between the molecules increases, the direct intermolecular interaction decreases. Consequently, the transmission probability of D and E decreases between the HOMO – LUMO gap and approaches the values that is expected classically (*T(E)* of A+B). This is due to a combination of two main effects. First, the HOMO energy level is moved up in energy as a result of the intermolecular interactions and secondly, the HOMO-LUMO energy gap decreases. The latter is due to the splitting of degenerate energy levels associated with the parallel molecules.

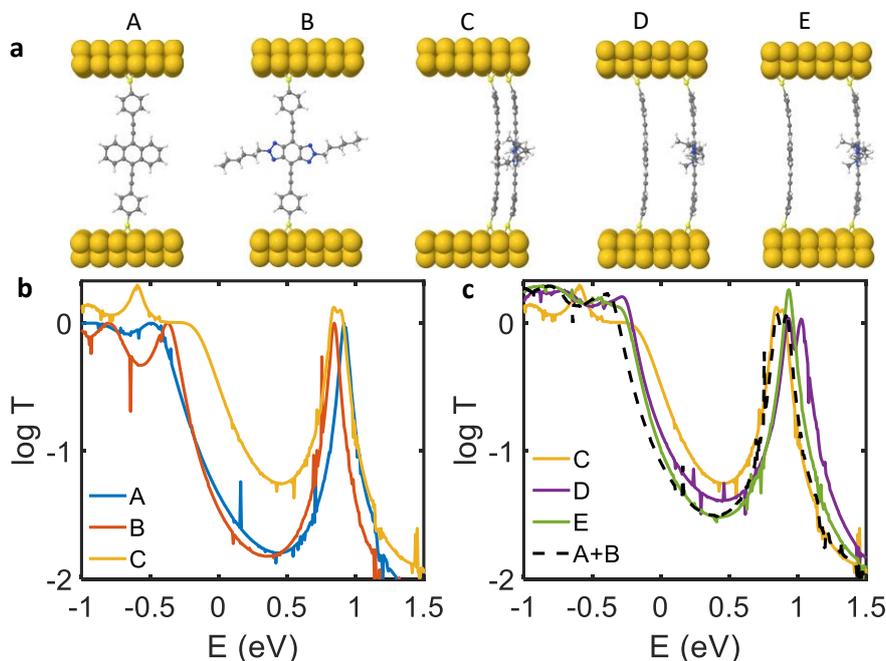

**Figure 2. Transport properties of molecular junctions.** (a) Molecular junctions formed by single molecule **1** and **2** (B and A) and dimer of **1** and **2** with different distance between them (C, D and E), (b) Transmission coefficients of A, B and C and (c) Transmission coefficients of D and E and comparison with C and A+B.

To further demonstrate the effect of intermolecular interaction on charge transport through molecular junctions, we built a simple tight-binding model as shown in Figure 3a. This includes two 2-levels systems with on-site energies *ε* coupled to each other through coupling integrals *γ*. They interact with each other through the intermolecular coupling *δ* and connected to the electrodes through couplings *α* and *β*. The parameters used for these calculations are shown in the SI. Figure 3b shows the changes in the energy levels of this two 2–level system as a function of changes in the intermolecular coupling δ. Clearly, the degenerate HOMO and LUMO levels split by increasing *δ*. As a result of this, the HOMO and LUMO energies are getting closer to each other if the intermolecular coupling is smaller than coupling between sites for each branch (*δ<γ*) and then move away from each other (*δ>γ*). The intermolecular coupling is expected to be smaller than internal coupling within the molecules in SAM and therefore the *δ<γ* regime is experimentally more accessible compared to the *δ<γ* regime. This is in agreement with our first-principle calculations using DFT where HOMO-LUMO gap shrinks as a result of interaction between molecules (configuration C in Figure 2a) and HOMO transport resonance move up in energy (Figure 2b).

In order to further demonstrate the effect of intermolecular interaction, we consider three scenarios. Figure 3c shows the transmission coefficient in the absence (case 1 where *δ=0*) and presence (case 2 where *δ≠0*) of the intermolecular interaction. As a result of intermolecular interactions, the HOMO energy level is moved up in energy and consequently, the HOMO resonance gets closer to the Fermi energy (*E=0eV*). Furthermore, the HOMO-LUMO energy gap decreases similar to what we obtained from the material specific DFT calculations. The molecular orbitals of dimer for each transport resonance are shown in the inset of Figure 3c. Interestingly, the self-energy vanishes for some of the resonances when $\alpha=\beta$ (Case 2) but restored by making the couplings to electrodes asymmetrical ($\alpha\neq\beta$, Case 3). As a result of this, two narrow resonances appear in $T(E)$ (orange line in Figure 3c). Since Seebeck coefficient is proportional to the slope of transmission coefficient, this sharp narrow resonance can be helpful to achieve further enhancement of Seebeck coefficient and power factor. This means that small disorder in the coupling to electrodes can even be advantageous for thermoelectricity.

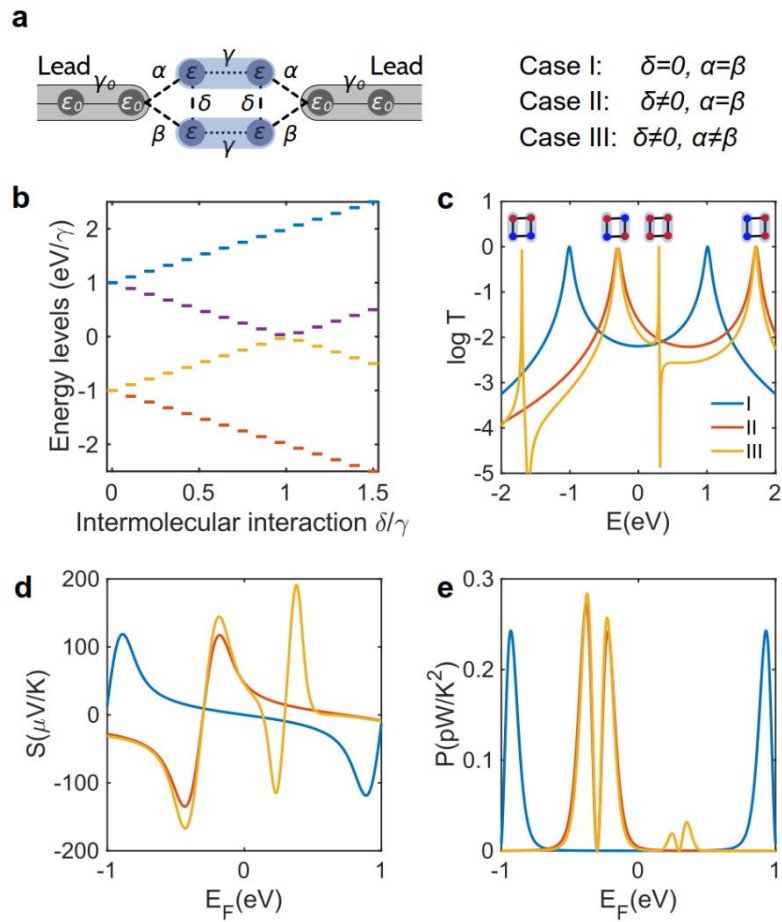

**Figure 3. Tight-binding model of a simple system including two parallel 2-levels branches.** (a) Two 2-levels system parallel to each other with intermolecular interaction $\delta$ connected to the electrodes with coupling ($\alpha$, $\beta$), (b) changes of the energy level of model system as a function of intermolecular interactions, transmission coefficient (c), room temperature Seebeck coefficient (d) and power factor (e) as a function of Fermi energy $E_F$ for structure shown in (a) with three different set of parameters Case I-III.

In order to investigate the effect of intermolecular coupling on thermoelectric properties of molecules, using the DFT transmission functions shown in Figure 2, we calculate room-temperature electrical conductance ($G$), Seebeck coefficient ($S$) and Power factor ($GS^2$) for structures A-C as shown in Figure 4a-c, respectively (see methods). As a result of intermolecular interaction, both $G$ and $S$ are enhanced simultaneously (Figures 4a,b). For example, the electrical conductance for dimer of **1** and **2** (structure C) is higher than both single molecule junctions A and B for all Fermi energies between the HOMO and LUMO (Figure 4a). The Seebeck coefficient also improves for energies around the DFT Fermi energy (shown by black dashed line in Figure 4). Figure 4d shows the calculated electrical conductance, Seebeck coefficient and power factor at room temperature and the Fermi energy shown by the black dashed line. Clearly, the power factor is enhanced by more than an order of magnitude in dimer C. This is because both the amplitude of transmission coefficient and its slop increases for dimer of **1** and **2** compared to that of single molecules **1** and **2**. Note that $G$ is proportional to the amplitude of $T(E)$ whereas $S$ is proportional to its slope[16].

Our simple tight-binding model of two parallel 2-levels system also shows similar behaviour as shown in Figure 3d,e. Since the HOMO resonance shifts up in energy and gets closer to the Fermi energy, the amplitude of $T(E)$ and its slop close to the Fermi energy increase leading to an increase in both $G$ and $S$ (Figure 3d). These combined leads to improvement of power factor (Figure 3e). In order to demonstrate that the improvement of thermoelectric power factor is a generic feature of dimers formed by placing conjugated molecules parallel to each other, we also investigated thermoelectric properties of dimers formed by two single molecules **2** with the anthracene core (as shown in Figure S5-6 of the SI). Our result shows similar behaviour to that of obtained for dimer of **1** and **2**. As shown in Figure S5-6, the power factor improves by more than an order of magnitude.

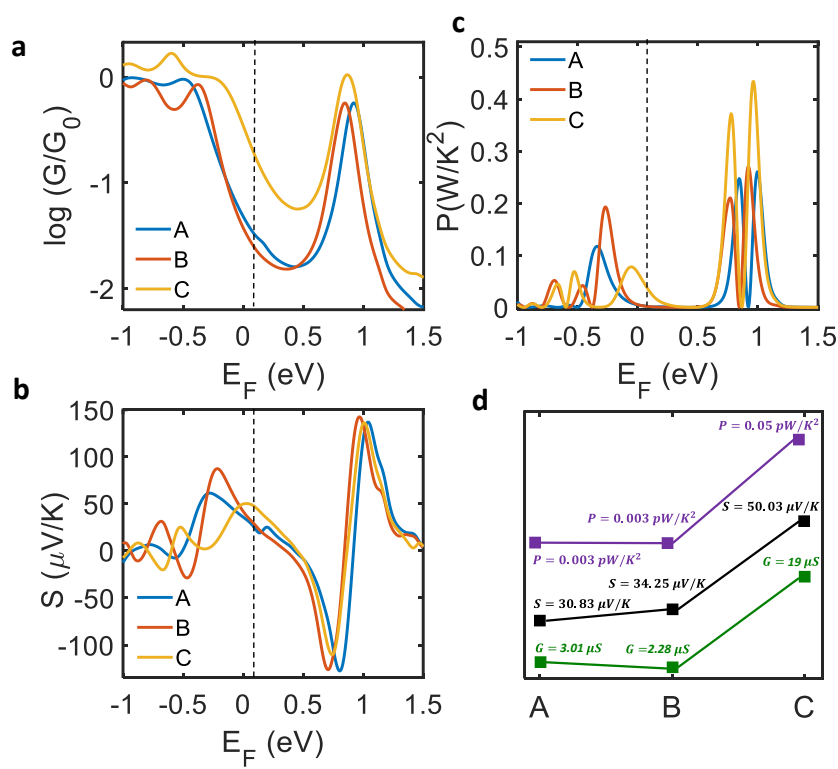

**Figure 4.** Thermoelectric properties of single and dimer molecules **1** and **2**. Room temperature electrical conductance (a), Seebeck coefficient (b) and power factor (c) versus electrodes Fermi energy for structures A-C shown in Figure 1a. (d) Comparison of room temperature electrical conductance , Seebeck coefficient and power factor at the Fermi energy (black dashed line).

It is worth mentioning that the enhancement of $G$ and $S$ simultaneously depends on how close a transport resonance can get near the Fermi energy. In the HOMO dominated molecular junctions, moving HOMO up in energy will enhance $G$ and $S$ simultaneously whereas in the LUMO dominated junctions, moving LUMO down will enhance $G$ and $S$. In addition to the intermolecular coupling, this also depends on the direction of charge transfer between the molecules. Therefore, it is important to identify both electron donating and deficient molecules to donate or withdraw charge from a molecular backbone of interest.

## Conclusion

In summary, we demonstrated that the intermolecular coupling between molecules leads to enhancement of thermoelectric properties of junctions formed by single molecules in parallel. We showed that electrical conductance and Seebeck coefficient enhance simultaneously as a result of intermolecular interaction between molecules with BBT and anthracene cores. These combined leads to enhancement of power factor. This is a result of two effects: due to intermolecular couplings, first, the energy levels move up in energy and secondly, the HOMO-LUMO gap decreases in dimers. Our result shows that intermolecular interaction can be used as a strategy to enhance thermoelectric efficiency further in tightly packed SAMs formed by single molecules.

## Computational methods

The optimized geometry and ground state Hamiltonian and overlap matrix elements of each structure (as shown in Figure 2a.) was self-consistently obtained using the SIESTA implementation[29] of density functional theory (DFT). SIESTA employs norm-conserving pseudo-potentials to account for the core electrons and linear combinations of atomic orbitals to construct the valence states. The generalized gradient approximations (GGA) of the exchange and correlation functional is used with PBE parameterization a double-ζ polarized (DZP) basis set, a real-space grid defined with an equivalent energy cut-off of 150 Ry. The geometry optimization for each structure is performed to the forces smaller than 10 meV/Å. We also performed calculations with the local density approximation (LDA) with CA parameterization and van der Waals (vdW) functional with DH using Siesta and get similar result to our calculations using GGA functional. The orbital calculations also were performed using Gaussian g16 code with tight convergence criteria and B3LYP hybrid functional that also shows similar trends to our Siesta calculations.

The mean-field Hamiltonian obtained from the converged DFT calculation was combined with GOLLUM[16,30] implementation of the non-equilibrium Green's function method to calculate the phase-coherent, elastic scattering properties of the each system consist of left gold (source) and right gold (drain) leads and the scattering region. The transmission coefficient $T(E)$ for electrons of energy $E$ (passing from the source to the drain) is calculated via the relation: $T(E) = Trace(\Gamma_R(E)G^R(E)\Gamma_L(E)G^{R\dagger}(E))$. In this expression, $\Gamma_{L,R}(E) = i\left(\sum_{L,R}(E) - \sum_{L,R}^{\dagger}(E)\right)$ describe the level broadening due to the coupling between left (L) and right (R) electrodes and the central scattering region, $\sum_{L,R}(E)$ are the retarded self-energies associated with this coupling and $G^R = (ES - H - \sum_L - \sum_R)^{-1}$ is the retarded Green's function.

The electrical conductance is then calculated using the Landauer formula $G(E_F, T) = G_0 L_0$, $L_n = \int_{-\infty}^{+\infty} dE\, (E - E_F)^n\, T(E)(-\partial f(E,T,E_F)/\partial E)$ and $f = (e^{(E-E_F)/k_B T} + 1)^{-1}$ is the Fermi-Dirac probability distribution function, $T$ is the temperature, $E_F$ is the Fermi energy, $G_0 = 2e^2/h$ is the conductance quantum, $e$ is electron charge and $h$ is the Planck's constant. The Seebeck coefficient is given by $S(E_F, T) = -\frac{L_1}{eTL_0}$ and Power factor is $P = GS^2$

## Conflicts of interest

There are no conflicts to declare.

## Acknowledgements

H.S. acknowledges the UKRI for Future Leaders Fellowship number MR/S015329/2. S.S acknowledges the Leverhulme Trust for Early Career Fellowship no. ECF-2018-375.